\documentclass[a4paper,french]{rnti}
\usepackage[T1]{fontenc}
\usepackage[latin1]{inputenc}
\usepackage{amsfonts}
\usepackage{mathtools}
\usepackage{amsmath}
\usepackage{url}

\usepackage{graphicx}

\usepackage[bookmarks=true, bookmarksnumbered=true, bookmarksopen=true,
		unicode=true, colorlinks=true,
		pagebackref=true,
		linkcolor=blue,citecolor=blue,filecolor=blue,urlcolor=blue
		]{hyperref}

\titrecourt{Une méthode pour caractériser les communautés des réseaux dynamiques à attributs}

\nomcourt{G. K. Orman \textit{et al}.}

\titre{Une méthode pour caractériser les communautés des réseaux dynamiques à attributs}

\auteur{Günce Keziban Orman\affil{1}\affilsep\affil{2}\,
        Vincent Labatut\affil{2}\,
        Marc Plantevit\affil{3}\,
        Jean-François Boulicaut\affil{1}}

\affiliation{
    \affil{1}Université de Lyon, CNRS, INSA-Lyon, LIRIS UMR5205, F-69621, France\\
          jean-francois.boulicaut@insa-lyon.fr, \\
    \affil{2}Université Galatasaray, Département d'informatique, Ortaköy 34349, Istanbul, Turquie  \\
          korman@gsu.edu.tr, vlabatut@gsu.edu.tr\\
     \affil{3} Université de Lyon, CNRS, Université Lyon 1, LIRIS UMR5205, F-69622, France \\
             marc.plantevit@liris.cnrs.fr\\

 }

\resume{%
De nombreux systèmes complexes sont étudiés via l'analyse de réseaux dits complexes ayant des propriétés topologiques typiques. Parmi celles-ci, les structures de communautés sont particulièrement étudiées. De nombreuses méthodes permettent de les détecter, y compris dans des réseaux contenant des attributs nodaux, des liens orientés ou évoluant dans le temps. La détection prend la forme d'une partition de l'ensemble des n\oe{}uds, qu'il faut ensuite caractériser relativement au système modélisé. Nous travaillons sur l'assistance à cette tâche de caractérisation. Nous proposons une représentation des réseaux sous la forme de séquences de descripteurs de n\oe{}uds, qui combinent les informations temporelles, les mesures topologiques, et les valeurs des attributs nodaux. Les communautés sont caractérisées au moyen des motifs séquentiels émergents les plus représentatifs issus de leurs n\oe{}uds. Ceci permet notamment la détection de comportements inhabituels au sein d'une communauté. Nous décrivons une étude empirique sur un réseau de collaboration scientifique.
}

\summary{%
Many complex systems are modeled through complex networks whose analysis reveals typical topological properties. Amongst those, the community structure is one of the most studied. Many methods are proposed to detect communities, not only in plain, but also in attributed, directed or even dynamic networks. A community structure takes the form of a partition of the node set, which must then be characterized relatively to the properties of the studied system. We propose a method to support such a characterization task. We define a sequence-based representation of networks, combining temporal information, topological measures, and nodal attributes. We then characterize communities using the most representative emerging sequential patterns of its nodes. This also allows detecting unusual behavior in a community. We describe an empirical study of a network of scientific collaborations.
}

\begin{document}
%\layout

% DEBUT DE L'ARTICLE
%
\section{Introduction}

Un \textit{réseau complexe} est la représentation d'un \textit{système complexe} sous forme de graphe. Ils sont devenus très populaires en tant qu'outil de modélisation durant la dernière décennie car ils permettent de mieux comprendre le fonctionnement et la dynamique de certains systèmes
. Un réseau complexe \textit{ordinaire} ne contient que des n\oe{}uds et les liens existant entre eux ; cependant il est possible de l'enrichir avec différents types de données : orientation et/ou poids  des liens, dimension temporelle, attributs associés aux n\oe{}uds ou aux liens, etc.  Cette souplesse a permis d'utiliser les réseaux complexes pour étudier les systèmes du monde réel dans de nombreux domaines : sociologie, physique, génétique, informatique, etc. \citep{no1}. 

La nature complexe des systèmes modélisés entraine la présence de propriétés topologiques non-triviales au sein des réseaux correspondants. Parmi celle-ci, la \textit{structure de communautés} est l'une des plus répandue et des plus étudiées. Informellement, on peut définir une communauté comme un groupe de n\oe{}uds densément interconnectés relativement au reste du réseau \citep{no1}. Cependant, dans la littérature, cette notion est formalisée de très nombreuses différentes façons \citep{no39}. Il existe en fait des centaines d'algorithmes destinés à détecter les structures de communautés, caractérisés par l'utilisation d'une définition et/ou d'un traitement différents. Certains sont basés sur la mesure de modularité, une mesure similarité entre n\oe{}uds,  le principe de compression des données, la notion de signification statistique, les mécanismes de diffusion de l'information, la percolation de cliques, etc. (cf. \citep{no39} pour une revue détaillée). La plupart des méthodes existantes traitent des réseaux ordinaires mais de nouvelles méthodes apparaissent pour analyser les réseaux les plus riches en exploitant les directions et poids des liens, puis le temps, et plus récemment les attributs des n\oe{}uds \citep{no11, no12,no13}. Ces dernières se concentrent sur la recherche de groupes de n\oe{}uds denses en termes de liens, et dont les attributs sont homogènes. Même si cela n'est pas toujours indiqué explicitement, ces méthodes exploitant les attributs s'appuient sur l'hypothèse que  les n\oe{}uds d'une même communauté doivent être similaires en termes d'attributs.
Bien que les algorithmes diffèrent en termes de nature des communautés détectées, de complexité algorithmique, de qualité du résultat et d'autres aspects \citep{no39}, leur production peut toujours être essentiellement décrite comme une liste de groupes de n\oe{}uds. Plus précisément, dans le cas de communautés mutuellement exclusives, il s'agit d'une partition de l'ensemble des n\oe{}uds. D'un point de vue applicatif, la question est alors de donner un sens à ces groupes relativement au système étudié. L'interprétation manuelle de petites communautés est possible, mais la méthode ne s'applique pas bien à de très grands réseaux. 

Seuls quelques travaux ont essayé de s'attaquer explicitement à ce problème. Dans \citep{no16}, les communautés sont caractérisées en comparant les distributions de plusieurs mesures purement topologiques. Dans \citep{no17} et \citep{no18}, les auteurs se concentrent sur les attributs nodaux, et identifient les plus représentatifs pour chaque communautés au moyen de divers outils statistiques classiques. Les méthodes de détection de communautés exploitant les attributs sont généralement aussi en mesure de donner ce type d'information, car ces attributs sont identifiés pendant le processus de détection.
Cependant, aucune des méthodes citées ne prend en compte toutes les données qu'un réseau riche peut contenir (attributs, topologie et dimension temporelle), qui plus est de manière systématique. Il existe donc un besoin pour un tel procédé, qui permettrait de caractériser les communautés des réseaux complexes riches. Dans ce travail, nous proposons une solution à ce problème, sous la forme d'une méthode d'analyse des réseaux attribués dynamiques. Pour cela, nous détectons les changements communs dans les mesures topologiques et les valeurs d'attribut sur une période de temps donnée. Plus précisément, nous cherchons à trouver les motifs séquentiels fréquents les plus représentatifs pour chaque communauté. Ces motifs peuvent ensuite être utilisées à la fois pour caractériser la communauté, et pour identifier ses anomalies, i.e. ses n\oe{}uds ayant un comportement non-standard. Les motifs fréquents représentent la tendance générale des n\oe{}uds dans la communauté considérée, alors que les anomalies peuvent correspondre à des n\oe{}uds ayant un rôle spécifique dans la communauté, ou situés à sa frontière. Nous illustrons notre méthode en l'appliquant à un réseau dynamique de co-auteurs extrait de la base de données bibliographiques DBLP\footnote{
%{\tt http://www.informatik.uni-trier.de/~ley/db/}
\url{http://www.informatik.uni-trier.de/~ley/db/}
}.

Notre première contribution est de considérer la caractérisation de communauté comme un problème spécifique, distinct de celui de la détection de la communauté. La méthode à appliquer doit être indépendante de la technique utilisée pour détecter les communautés, se fonder sur une approche systématique facilement reproductible, et être la plus  automatisée possible. Notre deuxième contribution est l'introduction d'une nouvelle représentation des réseaux attribués dynamiques. Elle prend la forme d'une base de données contenant des séquences de mesures topologiques, d'attributs nodaux et d'information communautaire, pour plusieurs tranches temporelles. Ce type de représentation avait précédemment été utilisé pour la représentation de données naturelles \citep{no19}, mais pas celle de graphes. Notre troisième contribution est la définition d'une méthode basée sur l'extraction sous contraintes de motifs séquentiels qui tire parti de cette représentation pour caractériser les communautés. Enfin, notre dernière contribution concerne une application à un réseau du monde réel. 
Dans la section suivante, nous donnons une description détaillée de notre méthode. Dans la section \ref{sec:resultats}, nous présentons nos résultats expérimentaux obtenus sur les données DBLP. La section \ref{sec:relatifs} décrit les travaux connexes, et la section \ref{sec:conclusion} présente les extensions possibles de notre travail.

\section{Méthode}
\label{sec:methode}
Un réseau dynamique  attribué est constituée de différentes tranches temporelles, chacune représentée par un sous-réseau distinct, contenant les liens entre les n\oe{}uds pour un intervalle de temps donné. Ces tranches temporelles sont séquentielles. Habituellement, les n\oe{}uds et leurs attributs sont les mêmes pour chaque tranche, tandis que les liens entre eux et les valeurs des attributs peuvent changer. Nous proposons de caractériser les communautés d'un réseau dynamique attribué en fonction de l'évolution commune des mesures topologiques et des attributs de leurs n\oe{}uds. Le processus que nous proposons inclut $4$ étapes. La première consiste à identifier une structure de communautés de référence. La seconde vise à créer la structure de données permettant une représentation séquentielle des mesures topologiques et des attributs des n\oe{}uds. Nous calculons d'abord les valeurs de toutes les mesures topologiques sélectionnées, puis nous discrétisons les valeurs des mesures et des attributs. Lors de la troisième étape, nous recherchons des motifs séquentiels fréquents et nous extrayons les n\oe{}uds qui supportent chaque motif. La quatrième étape consiste à choisir les motifs les plus représentatifs pour caractériser les communautés selon différents critères.

%\subsection{Détection des communautés}
\paragraph{Détection des communautés.}
Pour détecter comment les n\oe{}uds évoluent en fonction de l'appartenance communautaire, nous avons d'abord besoin d'une structure de communautés de référence. Il serait possible d'appliquer une méthode dynamique, cependant cela entraîne des complications dues aux fusions, séparations, disparitions et apparitions de communautés au cours du temps. Pour cette raison, dans ce premier travail, nous avons décidé d'utiliser des communautés statiques, détectés sur une version intégrée du réseau.

Nous créons d'abord un nouveau réseau en agrégeant tous les liens dans le temps. Un poids est attribué à chaque lien en fonction de son nombre d'occurrences, afin de représenter sa stabilité dans le temps. Nous appliquons ensuite un algorithme de détection de communautés classique pour identifier nos communautés statiques. À cette fin, nous avons sélectionné Louvain \citep{no5}, qui est une méthode reconnue pour obtenir des structures de communautés de bonne qualité. La complexité temporelle de cette algorithme est $ O(n\log n)$ où $n$ est  nombre de n\oe{}uds. La structure de communautés en résultant est utilisée dans le reste de notre analyse. Bien que les communautés soient statiques, les changements dans la structure du réseau seront néanmoins considérés lors du traitement des mesures topologiques des n\oe{}uds.

%\subsection{Constitution de la base de données}
%\label{subsect2.1}
\paragraph{Constitution de la base de données.}
La deuxième étape consiste à représenter le réseau d'une manière appropriée pour l'extraction des motifs séquentiels fréquents. Un réseau dynamique attribué $\mathcal{G} =\langle G_{1},\ldots,G_{\theta}\rangle$ est constitué d'une séquence de tranches temporelles, chacune prenant la forme d'un réseau attribué séparé $G_{j}$ $(1\leq j\leq\theta)$, représentant une période de temps donnée. Chaque réseau statique $G_{j}$ contient $n$ n\oe{}uds et correspond aux connexions entre ces n\oe{}uds pour la période $j$. Les n\oe{}uds et leurs attributs disponibles sont supposés être statiques, c'est à dire rester les mêmes pour tout $G_{j}$. Au contraire, les valeurs des attributs et la structure du réseau peuvent changer au cours du temps. Un descripteur de n\oe{}ud est soit un attribut nodal, soit une mesure topologique nodale. 

Nous avons sélectionné les mesures topologiques nodales les plus répandues (degré, transitivité locale), ainsi que certaines mesures permettant de caractériser les n\oe{}uds en termes de position dans leur communauté (degré interne, coefficient de participation et enchâssement). Le \textit{degré} $d$ d'un n\oe{}ud est son nombre total des voisins directs et sob \textit{degré interne} $d_{int}$ est le degré calculé dans sa seule communauté. La \textit{transitivité locale}, $T=\bigtriangleup/(d(d-1))$, d'un n\oe{}ud est la densité des triangles auxquels il appartient. Plus précisément, il s'agit du rapport entre le nombre de triangles auxquels le n\oe{}ud appartient effectivement ($\bigtriangleup$), et le nombre maximal de triangles auxquels il pourrait appartenir, étant donné son degré (dénominateur). Le \textit{degré interne normalisé} $z$ et le \textit{coefficient de participation} $P$ sont deux mesures définies par \cite{no20}. La première exprime combien le n\oe{}ud est connecté à sa propre communauté, relativement aux autres n\oe{}uds de sa communauté. Il s'agit du $z$-score de son degré interne, i.e. du nombre de voisins directs dans sa propre communauté. Un \textit{hub} au sens de \cite{no20} est un n\oe{}ud dont le degré interne est élevé relativement au autres n\oe{}uds de sa communauté. La seconde mesure caractérise la distribution des voisins du n\oe{}ud parmi toutes les communautés : $P=1-\sum_{c}(d_{c}/d)^{2}$, où $d_{c}$ est le nombre de liens entre le noeud considéré et la communauté $c$. Il s'agit plus précisément de mesurer l'hétérogénéité de cette distribution. On obtient une valeur proche de $1$ si tous les voisins sont répartis uniformément entre toutes les communautés, et $0$ s'ils sont tous réunis dans la même communauté. L'\textit{enchâssement}, $e=d_{int}/d$, évalue à quel point les voisins d'un n\oe{}ud appartiennent à sa propre communauté \citep{no16}. À la différence du degré interne normalisé, l'enchâssement est donc normalisé par rapport au n\oe{}ud, et non pas à sa communauté. Toutes ces mesures topologiques peuvent être calculées en un temps linéaire par rapport à $n$. %référence: http://jgaa.info/accepted/2005/SchankWagner2005.9.2.pdf

Soit $D=\{D_{1},D_{2},\ldots,D_{k}\}$ l'ensemble des descripteurs. Chacun peut prendre plusieurs valeurs discrètes, définies sur son \textit{domaine} $\mathcal{D}_i$ $(1\leq i\leq k)$. Toutes nos mesures topologiques sont à valeurs réelles, nous avons donc dû les discrétiser pour les adapter à cette définition. Un \textit{item} $I=\{(D_{i},u) \in D \times \mathcal{D}_{i} \}$ est un couple constitué d'un descripteur $D_{i}$ et d'une valeur $u$ de son domaine $\mathcal{D}_{i}$. Une séquence nodale est notée $\nu= \langle(l_{11},\ldots,l_{k1}),\ldots,$ $(l_{1 \theta},\ldots,l_{k \theta})\rangle$ où $l_{ij}$ est l'item du $i^{eme}$ descripteur pour la tranche temporelle $j$. Une base de données  $M$ est l'ensemble des tuples $(\eta, \nu_\eta ,C_{\eta})$ où $\eta$ est l'identifiant du n\oe{}ud, $\nu_{\eta}$   est sa séquence et $C_{\eta}$ est l'étiquette de sa communauté (tels que définis à l'étape 1). La création de $M$ revient à réaliser séquentiellement le calcul des mesures topologiques et la détection de communautés, donc sa complexité temporelle est aussi $ O(n\log n)$.

%\subsection{Fouille de motifs séquentiels émergents}\\
\paragraph{Fouille de motifs séquentiels émergents.} La troisième étape a pour but de trouver des motifs séquentiels émergents permettant de caractériser les communautés. Pour chaque communauté, nous recherchons les motifs séquentiels supportés par la majorité de ses membres. Avant de présenter la façon dont nous traitons la fouille, nous donnons quelques définitions nécessaires à la bonne compréhension de l'algorithme que nous avons utilisé.

Un \textit{itemset} $T$ est un sous-ensemble de $I$. Une \textit{séquence} $s= \langle t_{1},\ldots,t_{m}\rangle$ est une liste d'itemsets ordonnés dans le temps, où $m$ est la longueur de la séquence. Une séquence $\alpha=\langle a_{1},\ldots,a_{m} \rangle$ est une \textit{sous-séquence} d'une autre séquence $\beta=\langle b_{1},\ldots$
$,b_{n}\rangle$ ssi $\exists i_{1},i_{2},\ldots,i_{m}$ tels que $1\leq i_{1}<i_{2}<\ldots<i_{m}\leq n$ et $a_{1}\sqsubseteq b_{i1},a_{2}\sqsubseteq b_{i2}, \ldots,a_{m}\sqsubseteq b_{im}$ On dit également que $\beta$ est une \textit{super-séquence} de $\alpha$. La taille d'une communauté $C$ est le nombre total de n\oe{}uds qu'elle contient. Le support d'une séquence $s$ pour une communauté donnée $C$ est le rapport du nombre de n\oe{}uds supportant $s$ à la taille de $C$ : $sup(s,C)= |\{\eta\in C| s\ \sqsubseteq \nu_\eta\}| / |C|$. Le taux de croissance d'une séquence $s$ pour une communauté donnée $C$ est le rapport du support de $s$ dans $C$ au support de $s$ dans $\overline{C}$, où $\overline{C}$ est le complémentaire de $C$ dans le réseau (i.e. tous les n\oe{}uds du réseau sauf ceux de $C$) : $Gr(s,C)= sup(s,C) / sup(s,\overline{C})$. Le taux de croissance mesure l'émergence de la séquence $s$ dans la communauté $C$. Plus il est élevé, et plus la séquence $s$ est caractéristique de la communauté $C$.

Compte tenu d'un seuil de support minimal $min_{sup}$, un motif séquentiel fréquent est une séquence dont le support est supérieur ou égal à $min_{sup}$. Un motif séquentiel fréquent \textit{fermé} pour une communauté donnée est un motif séquentiel qui n'a pas de super-séquence pour le support minimal spécifié. Dans notre problème, nous caractérisons chaque communauté selon le motif séquentiel fréquent fermé. La méthode Clospan \citep{no21} a été définie pour identifier tous les motifs séquentiels fermés existants, pour un seuil donné, avec une complexité en $O(n^{2})$. Cependant, nous voulons que les motifs séquentiels identifiés soient représentatifs des communautés dans lesquelles ils sont identifiés. Il est donc nécessaire de prendre en compte le taux de croissance des motifs afin de tenir compte de leur émergence. Pour ce faire, nous appliquons la méthode de post-traitement définie dans \citep{no22} pour calculer les taux de croissance de séquences d'article classés. Au final, notre approche est donc la suivante : d'abord nous identifions les motifs séquentiels fermés d'une communauté pour un support minimum donné. Puis, nous calculons les supports de ces motifs pour le reste du réseau entier. Enfin, nous en déduisons le taux de croissance de chacun des motifs séquentiels fermés.

%\subsection{Sélection des motifs séquentiels}
\paragraph{Sélection des motifs séquentiels.}
Une fois que les motifs séquentiels ont été extraits pour chaque communauté comme expliqué à l'étape 3, nous devons sélectionner les plus représentatifs afin de caractériser la communauté. Pour cela, nous extrayons d'abord les n\oe{}uds supportant chaque motif. Dans un premier temps, nous avons décidé d'effectuer ce calcul de façon naive, ce qui entraine une complexité en $O(rn)$, où $r$ est le nombre de motifs. Mais il faut souligner que ce calcul pourrait être accéléré en l'intégrant à Clospan. On choisit ensuite les motifs selon deux approches distinctes : (1) motif dont le support est le plus élevé et (2) motif dont le taux de croissance est le plus élevé. Les résultats de l'étape 3 nous donnent directement le support et le taux de croissance. Cependant, dans certaines communautés, le motif de taux de croissance maximal ne couvre pas une partie significative des n\oe{}uds constituant la communauté. Il est alors nécessaire d'identifier d'autres motifs, dans un souci de représentativité. Nous sélectionnons alors le motif dont l'ensemble des n\oe{}uds le supportant est le plus différent possible de celui du premier motif, afin de couvrir au maximum la partie de la communauté ignorée par celui-ci. La distance entre les ensembles des n\oe{}uds supportant est calculée au moyen du coefficient de Jaccard. En cas d'égalité, nous utilisons le taux de croissance comme critère secondaire. Si la couverture totale est toujours insuffisante, on réitère en sélectionnant d'autres motifs selon les mêmes modalités. Cette itération se poursuit jusqu'à ce qu'au plus cinq  n\oe{}uds ne soient pas couverts. Nous qualifions de \textit{déviant} ces n\oe{}uds ne supportant aucun motif représentatif de la communauté. Comme nous devons comparer chaque paire de motifs pour chaque communauté, le temps nécessaire à la sélection des motifs représentatifs est de l'ordre de $O(r^{2}/p)$ où $p$ est le nombre de communautés. La complexité totale de notre méthode est donc $O(n^{2}+r^{2}/p)$. Il faut noter que $r$ dépend grandement des nombres de tranches temporelles $ \theta$ et d'items $|I|$.

\section{Résultats}
\label{sec:resultats}
Nous présentons les expérimentations réalisées sur des données réelles. Nous avons choisi de traiter le réseau dynamique attribué de co-auteurs décrit dans \citep{no23} et extrait de la base de données DBLP. Chacun des $2145$ n\oe{}uds représente un auteur. Deux n\oe{}uds sont reliés si les auteurs correspondants ont publié au moins un article ensemble. Chaque tranche temporelle correspond à une période de $5$ ans. Il y a au total $10$ tranches temporelles allant de $1990$ à $2012$. Les périodes consécutives ont un chevauchement de $3$ ans pour des raisons de stabilité. Pour chaque auteur, à chaque tranche temporelle, la base de données fournit le nombre de publications pour $43$ conférences et journaux. Nous utilisons ces informations pour définir les $43$ attributs nodaux correspondants, et nous en rajoutons deux : le nombre total de publications dans les conférences et dans les journaux. On a donc un total de $45$ attributs. Nos descripteurs sont ces attributs, ainsi que les mesures topologiques décrites à la section \ref{sec:methode}.

Les mesures topologiques sont discrétisées différemment, en fonction de leur nature. Pour le degré, nous utilisons les seuils $3$, $10$ et $30$. Pour la transitivité, qui est définie sur $[0;1]$, il s'agit de $0,35$, $0,5$ et $0,7$. Pour l'enchâssement, lui aussi défini sur $[0;1]$, les intervalles sont $0,3$ et $0,7$. Ces intervalles ont été déterminés de manière à tenir compte des distributions de ces mesures sur l'ensemble des n\oe{}uds et des tranches temporelles : les différents seuils correspondent aux zones de faible densité. Pour les mesures de \cite{no20}, nous utilisons les seuils définis dans l'article original, c'est-à-dire : $2,5$ pour $z$ et $0,05$, $0,6$, et $0,8$ pour $P$. Le seuil utilisé pour $z$ permet de distinguer les hubs  $(z>2,5)$ des non-hubs $(z\leq2,5)$ communautaires. Pour les deux attributs relatifs aux nombres de publications dans des conférences et revues, nous considérons les valeurs $1, 2, 3, 4$ et $>5$. Pour le nombre total de publications, nous avons identifié les seuils $5$, $10$, $20$ et $50$ comme étant les plus pertinents.

\begin{table}[ht]
 \begin{center}
   \tabcolsep = 3\tabcolsep
   \begin{tabular}{rrrr}
   \hline\hline
   Communauté & Taille & Longueur du motif & Support\\
   \hline
   38 & 335       & 2     & 0,99   \\
   40 & 43        & 8     & 0,97   \\
   42 & 109       & 5     & 1,00   \\
   45 & 227       & 3     & 1,00   \\
   55 & 39       & 7     & 1,00   \\
   61 & 204       & 3     & 1,00   \\
   75 & 140       & 4     & 0,99   \\
   77 & 41       & 7     & 1,00   \\
   86 & 111       & 3     & 1,00   \\
   98 & 113       & 5     & 1,00   \\
   106 & 134       & 5     & 1,00   \\
   115 & 125       & 1     & 1,00   \\
   125 & 79       & 3     & 1,00   \\
   \hline
   \end{tabular}
\caption{Longueur des motifs les plus supportés pour une sélection de communautés.} \label{tab_exemple}
 \end{center}
\end{table}

L'algorithme Louvain identifie $127$ communautés dans le réseau pondéré global, pour une modularité de $0,59$. Cette valeur signifie que ce réseau est hautement modulaire. $96$ de ces communautés ne contiennent qu'un seul n\oe{}ud. Parmi les communautés restantes, $17$ contiennent plus de $10$ n\oe{}uds, la plus grande en ayant $335$. Nous recherchons ensuite les motifs séquentiels de ces communautés, pour un support minimum de $0,3$. Il ne nous a pas été possible de descendre en dessous de cette valeur en raison du coût spatial de l'algorithme Clospan utilisé. Pour chacune des communautés dont la taille est supérieure à $40$, nous détections plus de $5000$ motifs, dont la plupart ne comprennent que des mesures topologiques.

Les motifs les plus supportés sont toujours une séquence de $z<2,5$ pour toutes les communautés, avec des longueurs variables. Cela signifie que la majorité des n\oe{}uds de chaque communauté ont un rôle de non-hub communautaire. Pour mémoire, Amaral \& Guimerà définissent un hub communautaire comme un n\oe{}ud dont le degré interne est largement supérieur au degré interne moyen de sa communauté. Ainsi, le motif détecté signifie que la majorité des n\oe{}uds ont un degré interne relativement faible (ce qui est attendu), et ce durablement (ce qui ne l'est pas forcément). Bien que ce type de motif soit présent dans toutes les communautés, on peut faire une distinction en considérant la longueur de la séquence, qui mesure cette durée. Dans le Tableau \ref{tab_exemple}, nous listons la longueur des motifs séquentiels les plus supportés, en indiquant leur communauté, la taille de celle-ci et la valeur de support du motif. Les communautés dont les tailles sont comprises entre $39$ et $45$ (communautés $40$, $55$ et $77$ ) obtiennent de longues séquences (resp. $8, 7$ et $7$ ). Surtout, les supports pour les communautés $55$ et $77$ atteignent même la valeur maximale de $1$. Cela signifie que, dans ces communautés, aucun hub n'existe ; ou bien si un hub apparait, il disparait rapidement. Cette observation est particulièrement intéressante, et traduit l'absence d'un meneur communautaire qui structurerait la communauté par ses connexions multiples. Pour la communauté $115$, la longueur de la séquence est $1$ et sa valeur de support est également $1$. Cela signifie que tous les n\oe{}uds de cette communauté ont tenu le rôle de non-hub simultanément au moins une fois au cours du temps, mais qu'il existe des hubs pour le reste des tranches temporelles. Pour les communautés $38$, $40$ et $75$, le support est inférieur à $1$, ce qui signifie que si une écrasante majorité de n\oe{}uds tient le rôle de non-hub pour de longue durées, en revanche un petit nombre de n\oe{}uds occupe la place de hub, éventuellement par intermittence. Nous extrayons les auteurs qui ne suivent pas les motifs les plus supportés  pour ces trois dernières communautés. Pour la communauté $38$, il s'agit de \textit{Philip S. Yu}, \textit{Jiawei Han} et \textit{Beng C. Ooi}. Comme supposé, ces n\oe{}uds ont un nombre de connexions remarquablement élevé à l'intérieur de leurs communautés, et les auteurs qu'ils représentent ont effectivement des rôles de meneurs dans leur domaine. Une analyse plus approfondie des données montre également qu'ils publient un total de plus de $10$ articles par tranche temporelle. En outre, ils n'occupent jamais de rôle non-hub. Les n\oe{}uds déviants pour les communautés $40$ et $75$ sont respectivement \textit{Hans-Peter Kriegel} et \textit{Divesh Srivastava}. Là aussi, il s'agit d'auteurs importants dans leur communauté. Les séquences qui les caractérisent confirment qu'ils sont productifs et n'occupent jamais le rôle non-hub au fil du temps. 

Pour les communautés dont les tailles sont comprises entre $39$ et $45$, nous ne constatons aucune motif émergent contenant une conférence ou revue. Les motifs les plus émergents  ont un taux de croissance supérieur à $1,79$, ce qui signifie qu'il n'existe pas de motif séquentiel très distinctif pour ces communautés. Pour la plupart des grandes communautés, le motif le plus émergent inclut une conférence ou une revue spécifique, ce qui peut s'interpréter en termes de thématique de la communauté. Les autres descripteurs constituant le motif sont des mesures topologiques. Comme pour les motifs les plus supportés, l'item $z<2,5$ apparait le plus souvent dans les motifs détectés. Cependant, ces motifs les plus émergents ne permettent pas de couvrir la majorité des n\oe{}uds des communautés concernées. C'est pourquoi, comme nous l'avons expliqué dans la section \ref{sec:methode}, nous recherchons des motifs supplémentaires en minimisant l'intersection de leurs supports. Ces motifs sont généralement constitués de mesures topologiques, et n'ont pas un taux de croissance très élevé. Dans la suite, nous nous concentrons sur les communautés donnant les résultats les plus intéressants. Pour chacune, nous décrivons le motif le plus émergent et nous présentons les n\oe{}uds déviants, qui ne supportent ni le motif le plus émergent ni les motifs supplémentaires. Chaque motif est représenté formellement entre crochets, comme une séquence d'itemsets, eux-mêmes représentés entre parenthèses.

Pour la communauté $61$, le motif le plus émergent est \textit{<(ICML  PUB. NUM = 1) (DEGRE 3-10,  Z < 2,5)>} avec un taux de croissance $3,52$ et support $0,30$. Ce motif fait référence aux auteurs qui sont publiés une fois dans ICML, après quoi leur degré se stabilise entre $3$ et $10$ et ils occupent tous le rôle de non-hubs. Nous extrayons $7$ motifs supplémentaires afin de couvrir tous les n\oe{}uds de la communauté. Parmi eux, les plus intéressants sont \textit{<(Z < 2,5) (Z < 2,5) (Z < 2,5, TOTALE  CONF PUB. NUM  1-5) (AAAI  PUB. NUM = 1)>} avec un taux de croissance de $1,69$ et un support de $0,30$ et \textit{<(PART. COEFF 0.05-0.6, CIKM PUB. NUM = 1)>} avec un taux de croissance de $1,40$ et un support de $0,30$. Le premier  motif se réfère à des n\oe{}uds qui restent des n\oe{}uds non-hubs pendant un certain temps, puis commencent à publier dans des conférences, avant de publier à AAAI tout en perdant leur statut de non-hub (sans pour autant devenir massivement des hubs). Le second n'a pas de dimension temporelle, mais nous montre l'existence ponctuelle de n\oe{}uds publiant à CIKM tout en occupant une position périphérique dans leur communauté, i.e., en étant significativement connecté à d'autres communautés.  Les n\oe{}uds déviants pour cette communauté sont \textit{Alex A. Freitas, Claire Cardie , Edwin P. D. Pednault}. Parmi ces auteurs, Alex A. Freitas ne publie  pas pendant les $8$ premières tranches temporelles, puis commence à publier de manière très active, à différentes conférences telles qu'ICML ou AAAI, et dans des journaux. Pour les deux autres auteurs, tandis que Claire Cardie publie régulièrement à ICML au cours des $6$ premières tranches temporelles, Edwin P. D. Pednault ne publie ni à ICML, ni AAAI ou CIKM .

Le motif \textit{<(PODS PUB. NUM = 1)>} est le plus émergent dans la communauté $75$. Son taux de croissance est $3,59$ et son support $0,40$.  Ce motif montre que $\%40$ des auteurs de cette communauté publie au moins une fois à PODS et ce groupe est émergent par rapport au reste du réseaux pour cette communauté. Quatre motifs supplémentaires sont nécessaires pour couvrir le reste de la communauté. Ces motifs se réfèrent à des n\oe{}uds occupant des positions de non-hub périphérique, et dont la transitivité est très élevée. Les n\oe{}uds déviants sont \textit{Ninghui Li, Li Feifei} et \textit{Abdullah Mueen}, qui ont la particularité de ne jamais publier dans PODS. Le motif le plus émergent de la communauté $106$ est \textit{<(Z < 2,5) (Z < 2,5) (Z < 2,5) (Z < 2,5) (Z < 2,5) (PART. COEFF 0.05-0.6, KDD PUB. NUM=1)>} avec un taux de croissance $2,87$ et support $0,40$. Ce motif fait référence à des n\oe{}uds durablement non-hubs, qui deviennent périphériques tout en publiant à KDD.   Nous identifions $4$ motifs supplémentaires pour finir de couvrir la communauté. Ceux-ci concernent les n\oe{}uds à la fois ultrapériphériques et bien intégrés à leur propre communauté. Les n\oe{}uds déviants sont \textit{Stan Matwin}, qui publie plus d'un article par tranche temporelle à KDD, et qui n'est pas durablement non-hub ; et \textit{Hua-Jun Zeng} qui n'a jamais publié dans KDD. Ce dernier est aussi caractérisé par un fort accroissement du nombre d'articles produits, alors qu'il ne publie pas pendant les $5$ premières tranches temporelles. 

Le motif le plus émergent de la communauté $45$ est \textit{<(VLDB PUB. NUM=3) (DEGRE 3-10 Z < 2,5)>} avec un taux de croissance de $6,40$ et un support de $0,30$. Cette séquence nous dit qu'il y a un groupe remarquable d'auteurs qui ont publié $3$ fois en conférence VLDB, puis dont le degré s'est stabilisé entre $3$ et $10$ et qui ont occupé des positions non-hubs dans leur communauté.  Nous avons dû identifier $6$ autres motifs pour couvrir le reste de la communauté. L'un d'eux est \textit{<(Z < 2,5, TOTAL CONF. PUB. NUM  1-5) (Z < 2,5, EMBED 0.3-0.7, ICDE PUB. NUM= 1)>}, avec un taux de croissance de $2,30$ et un support de $0,30$. Ce motif couvre les n\oe{}uds non-hubs qui ont publié entre $1$ et $5$ fois dans des conférences, puis dont l'enchâssement s'est stabilisé à une valeur relativement élevée, tout en gardant leur position de non-hub et en publiant à ICDE . Les n\oe{}uds déviants sont \textit{Ingmar Weber} et \textit{Anastasia Ailamaki}, qui ne publient pas  pour les sept premières tranches temporelles, puis deviennent de plus en plus productifs au cours des trois dernières tranches.

Pour résumer nos observations, le motif le plus émergent dans à peu près toutes les  communautés comprend habituellement le fait d'être non-hub et d'avoir un petit nombre de publications dans divers conférence ou journaux. En fonction des quelques conférences ou journaux apparaissant dans ces motifs, il est possible de déduire le thème principal des communautés. Pour certaines communautés, cependant, les motifs séquentiels émergents sont purement topologiques (pas d'attributs). On peut alors supposer que les membres de ces communautés ne publient pas de façon suffisamment homogène pour que cela transparaisse dans les motifs. Une autre raison peut être simplement que les membres de la communauté sont reliés pour des raisons autres que thématiques, auquel cas cela n'apparait pas dans les attributs sélectionnés pour notre étude. En ce qui concerne les n\oe{}uds déviants, on peut distinguer différents types de profils. Certains peuvent correspondre à des auteurs dont la thématique principale est différente de celle de la communauté dans laquelle ils ont été placés. Dans certains cas, nous avons détecté des auteurs qui avaient visiblement changé de thématique, ou bien qui débutaient dans une thématique donnée. Il peut également s'agir d'auteurs actifs dans un autre domaine, dont les conférences et journaux ne font pas partie de ceux retenus dans les données que nous avons considérées ici. Un autre profil est celui du chercheur en train de monter en charge, et dont la position communautaire et le nombre de publications sont en train d'évoluer de conjointement.

\section{Travaux connexes}
\label{sec:relatifs}
Il existe de nombreuses méthodes de détection de communautés dans les réseaux complexes ordinaires (i.e., ne contenant que des n\oe{}uds et des liens), basées uniquement sur l'information topologique. Une revue détaillée de ces méthodes est donnée dans \citep{no39}. Dans le cas des réseaux plus riches, contenant des attributs nodaux, une communauté est essentiellement définie comme un groupe de n\oe{}uds à la fois densément interconnectés, et similaires en termes d'attributs. Certaines méthodes tentent d'estimer des partitions optimisant directement ces deux critères simultanément \citep{no24}. D'autres transforment encodent d'abord les attributs sous forme d'information structurelle, et détectent ensuite les communautés de façon plus classique \citep{no11, no13}. Il existe également des méthodes de recherche de motifs (ici de petits sous-graphes densément connectés tels que des cliques) possédant des attributs homogènes \citep{no29, no30, no32}. 

Il existe aussi des travaux qui se concentrent sur la détection de communautés dans les réseaux dynamiques \citep{no34}. Leur but principal est l'observation de l'évolution de la structure de communautés, à travers des évènements tels que la formation, la dissolution, la croissance, la diminution et la fusion. Pour des réseaux dynamique attribuées, il existe des méthodes pour trouver les groupes de n\oe{}uds structurellement similaires dont les attributs changement de la même façon au cours du temps \citep{no23}. Ces méthodes ne cherchent pas des communautés. L'objectif de ces travaux est la découverte de motifs intéressants en tenant compte de la structure des liens et des attributs pour les réseaux dynamiques attribuées. Pour regrouper les n\oe{}uds, ces méthodes utilisent des contraintes, comme par exemple une distance limite entre n\oe{}uds de la même motif, ou la nécessité de partager les mêmes voisins. 

Bien qu'il existe de nombreuses techniques différentes pour identifier les structures de communautés, il y a peu d'auteurs qui travaillent à caractériser les communautés obtenues. Dans \citep{no16}, les auteurs comparent les distributions de certaines mesures topologiques afin de comprendre la forme générale des communautés, et tentent de les caractériser en fonction du type de système modélisé (biologique, informatique, social, etc.). Ils n'utilisent pas d'attributs nodaux et ne traitent que des réseaux statiques. Dans \citep{no17}, les auteurs proposent une méthode statistique inspirée d'approches utilisées en génétique, pour caractériser les communautés en termes d'attributs surexprimés. Cependant, cette étude n'a pas recours à des mesures topologiques et se limite également aux réseaux statiques. Dans \citep{no18}, les auteurs interprètent les communautés d'un réseau attribué social. Ils utilisent la régression statistique et l'analyse discriminante pour identifier les valeurs des attributs les plus caractéristiques de chaque communauté. Cependant, ici encore, le réseau est statique et l'information structurelle n'est pas exploitée.

Enfin, le travail présenté dans \citep{no36} n'est pas directement concerné par la détection ou la caractérisation de communautés, mais est néanmoins lié à notre approche. \citet{no36} ont introduit l'idée d'utiliser les mesures topologiques et les attributs nodaux ensemble dans une perspective d'exploration de données. Leur but était de trouver des motifs reflétant la covariation de mesures topologiques et d'attributs sur l'ensemble du réseau. Ils se sont concentrés sur des réseaux statiques et n'ont pas considéré de structures communautaires.

\section{Conclusion}
\label{sec:conclusion}
Nous traitons le problème de la caractérisation des communautés dans des réseaux complexes dynamiques et attribués. Nous proposons une nouvelle représentation de l'information encodée dans le réseau, permettant de stocker simultanément l'information topologique, les attributs nodaux et la dimension temporelle. Nous utilisons cette représentation pour effectuer une fouille de motifs séquentiels fréquents. Chaque communauté peut ensuite être caractérisée par ses motifs les plus distinctifs. Nous tirons également parti des motifs pour détecter et caractériser les n\oe{}uds déviants dans chaque communauté. Nous appliquons notre méthode à un réseau de collaboration scientifique construit à partir des données publiques de la base DBLP. Les résultats montrent que notre méthode est capable de caractériser les communautés en fonction, notamment, de leur thématique. Les n\oe{}uds déviants identifiés correspondent à différents types de profils, tels que des chefs de file communautaires, des chercheurs émergents, ou d'autres en train de changer de thématique de recherche.

A notre connaissance, il s'agit de la première formulation de la caractérisation de communautés comme un problème de fouille de données. Notre but était de surmonter les limitations des rares travaux existants \citep{no16, no17, no18} en proposant une approche systématique, tenant compte à la fois de la structure, des attributs nodaux et du temps. La représentation des données que nous utilisons n'avait jamais été appliquée au traitement de graphes. Le processus proposé pour extraire les motifs les plus pertinents en se basant sur une recherche de motifs séquentiels sous contraintes est original et nous avons montré la richesse des interprétations apportées sur un cas réel de réseau. Certaines étapes de l'implémentation présentée ici étaient relativement naives, nous comptons les améliorer afin de réduire le temps nécessaire au calcul. En particulier, l'identification des ensembles de support pourrait être intégrée à Clospan.

Afin de restreindre la complexité conceptuelle de cette première approche, nous avons volontairement limité notre méthode d'analyse en ne considérant pas l'évolution des communautés au cours du temps. Dans des travaux ultérieurs, nous envisageons d'appliquer un algorithme de détection de communautés approprié en insérant cette information dans la base de données utilisée pour la fouille de motifs. Nous comptons également appliquer notre méthode d'analyse à d'autres types de réseaux pour explorer ses capacités de caractérisation. Comme autre perspective, nous pouvons aussi mieux exploiter nos représentations de réseaux attribués dynamiques. Ici, nous nous sommes seulement intéressés à l'extraction de séquences fréquentes. Cependant, notre représentation des données du réseau peut également être utilisée pour interroger les n\oe{}uds selon certaines mesures ou attributs topologiques spécifiques. Ainsi, dans nos expériences, nous avons vu qu'il y avait beaucoup de n\oe{}uds qui n'appartiennent pas à une communauté. Il serait intéressant de bien regarder ces n\oe{}uds et mieux comprendre en quoi ils sont différents des autres pour, par exemple, formuler des hypothèses sur leur isolement.

\bibliographystyle{rnti}
\bibliography{orman2014}

\appendix
\Fr

\end{document}